\pdfoutput=1
\documentclass[english]{article}
\usepackage{geometry}
\usepackage{babel}
\usepackage{xcolor}
\usepackage{graphicx}
\usepackage{amsmath,amsfonts}
\usepackage{xcolor}
\usepackage{graphicx,indentfirst,subfigure,epsfig}
 \usepackage{varioref}
 \usepackage{wrapfig}
 \usepackage{subfigure}
 \usepackage{subfigmat}
 \usepackage{amsthm}
\usepackage{bm}

  \def\clap#1{\hbox to 0pt{\hss#1\hss}}

\providecommand{\mat}[1]{\bm{#1}}%
\renewcommand{\vec}[1]{\mathbf{#1}}

\providecommand{\mA}{\ensuremath{\mat{A}}}
\providecommand{\mB}{\ensuremath{\mat{B}}}

\providecommand{\mF}{\ensuremath{\mat{F}}}

\providecommand{\mH}{\ensuremath{\mat{H}}}
\providecommand{\mI}{\ensuremath{\mat{I}}}

\providecommand{\mP}{\ensuremath{\mat{P}}}

\providecommand{\mR}{\ensuremath{\mat{R}}}

\providecommand{\mU}{\ensuremath{\mat{U}}}

\providecommand{\mX}{\ensuremath{\mat{X}}}

\providecommand{\va}{\ensuremath{\vec{a}}}

\providecommand{\vh}{\ensuremath{\vec{h}}}

\providecommand{\vn}{\ensuremath{\vec{n}}}

\providecommand{\vv}{\ensuremath{\vec{v}}}

\providecommand{\vz}{\ensuremath{\vec{z}}}

\usepackage{hyperref}
\hypersetup{colorlinks, citecolor=red, linkcolor=blue, urlcolor=red, breaklinks=true}
\usepackage{scrextend}
\usepackage[T1]{fontenc}
\usepackage[utf8]{inputenc}
\usepackage{tgtermes}
\usepackage{nomencl}
\usepackage{algorithm}
\usepackage{blkarray}
\usepackage{algorithmic}
\usepackage{booktabs}
\usepackage{longtable}
\usepackage{lineno}
\makenomenclature
\usepackage{soul}
\begin{document}
\large
\title{\textbf{Spatial Flow-Field Approximation Using \\Few Thermodynamic Measurements \\ Part II: Uncertainty Assessments}}
\author{Pranay Seshadri\footnote{Address all correspondence to \texttt{ps583@cam.ac.uk}}$\;^{\; \star \dagger}$, Andrew Duncan$^{\smallint \dagger}$, Duncan Simpson$^{\ddagger}$, \\George Thorne$^{\ddagger}$, Geoffrey Parks$^{\star}$\\ \vspace{0.3 cm} \\ $^{\star}$Department of Engineering, University of Cambridge, Cambridge, U. K., \\ $^{\dagger}$Data-Centric Engineering, The Alan Turing Institute, London, U. K., \\ $^{\smallint}$Department of Mathematics, Imperial College, London, U. K., \\$^{\ddagger}$Civil Aerospace, Rolls-Royce plc., Derby, U. K.}
\date{}
\maketitle{}
\begin{abstract}
In this second part of our two-part paper, we provide a detailed, frequentist framework for propagating uncertainties within our multivariate linear least squares model. This permits us to quantify the impact of uncertainties in thermodynamic measurements---arising from calibrations and the data acquisition system---and the correlations therein, along with uncertainties in probe positions. We show how the former has a much larger effect (relatively) than uncertainties in probe placement. 

We use this non-deterministic framework to demonstrate why the well-worn metric for assessing spatial sampling uncertainty falls short of providing an accurate characterization of the effect of a few spatial measurements. In other words, it does not accurately describe the uncertainty associated with sampling a non-uniform pattern with a few circumferentially scattered rakes. To this end, we argue that our data-centric framework can offer a more rigorous characterization of this uncertainty. Our paper proposes two new uncertainty metrics: one for characterizing \emph{spatial sampling uncertainty} and another for capturing the impact of \emph{measurement imprecision} in individual probes. These metrics are rigorously derived in our paper and their ease in computation permits them to be widely adopted by the turbomachinery community for carrying out uncertainty assessments.

%
%
%

\end{abstract}

\section{Introduction and Motivation}
\label{sec:intro}
In the second part of this two-part paper, we focus on the non-deterministic aspects of a spatial data-driven thermodynamic model. Recall, in \cite{seshadri2019a} we developed a 2D multilinear least squares analytical framework for approximating the temperature (or pressure) at an axial plane in an engine, given measurements from a few rakes. We made the assumption that the measurements were deterministic and that the probe positions were precisely known. Moreover, we showed how the averages computed using our model could offer more accurate and robust area average values compared to currently, widely-adopted, averaging practices. In this paper, we study how an uncertainty in both measurements and probe positions can affect our model and its computed averages. Our paper, however, has far greater scope than our model and its uncertainties, delving into the very definition of the experimental uncertainties and how they are aggregated. At the outset we remark that \emph{experimental uncertainty quantification} has been the subject of considerable work. Well-established guides such as the International Organization for Standardization (ISO) guide to the expression of uncertainty in measurements  \cite{iso200898}, the American Society of Mechanical Engineers (ASME) performance test codes (PTC) 19.1 \cite{dieck2005test}, and the European Accreditation guide \cite{ea2013} are industry standards used around the world for propagating experimental uncertainties; other guides include  \cite{abernathy1973handbook, dieck2007measurement}. It will be worthwhile to briefly outline the central ideas presented in these guides.

The ISO GUM \cite{iso200898} categorizes uncertainties into two types: \emph{type A} and \emph{type B}. Type A uncertainties are characterized and evaluated by statistical approaches, e.g. sampling, whereas type B uncertainties are assessed by non-statistical techniques, e.g. engineering judgment and historical trends. There are parallels between this division of uncertainties and the aleatory and epistemic description of uncertainties, the latter two being common parlance within the field of uncertainty quantification (UQ) \cite{smith2013uncertainty}. The ASME PTC 19.1 \cite{dieck2005test} splits uncertainties into \emph{systematic} and \emph{random} components. The systematic uncertainty can be interpreted as a bias in a measurement, while the random component can be associated with the precision of the measurement device. The ISO standard makes the argument that such a characterization may result in ambiguity as a random component of uncertainty in the measurement of one quantity can become a systematic uncertainty in the measurement of another quantity (see page 18 in \cite{iso200898}). Apart from this description of uncertainties, the techniques for propagating the uncertainties (where they can be propagated) remain similar; they are detailed below. 
\begin{itemize}
\item \underline{Law of propagation of uncertainty}: A first-order Taylor series analysis which requires details on individual variances and a functional form (see page 56 in \cite{iso200898}). It can also propagate the effect of correlations. A simplification of this law is the \emph{root sum square method}. As the name states, one simply takes the square root of the sum of the squares of the uncertainties (expressed as 95$\%$ standard deviations). In the ASME standard, the systematic and random uncertainties are aggregated separately using the same formulation (see page 162 in \cite{abernethy1985asme}). 
\item \underline{Monte Carlo}: Numerical estimates of uncertainties can be determined via Monte Carlo simulations; this is particularly useful in models where closed analytical representations of the uncertainties are not available. Monte Carlo samples and its variants e.g., importance sampling, rejection sampling (see \cite{mcbook}) are available for both cases where the inputs are independent and correlated. 
\end{itemize}
The challenge, however, is seldom which of the above strategies to adopt, but rather the identification and characterization of the uncertainty itself. One of our interests in this paper is in propagating uncertainties associated with sampling. The interpretation of finite spatial sampling as a source of error is duly noted in Pianko and Wazelt \cite{pianko1983propulsion}, where the authors state that \emph{``[sampling uncertainty] relates to the measurement property [e.g., temperature] distortion pattern and the number of flow measurements for a given spatial area''}. Historically (see 8.1.4.4.3 in \cite{saravanmuttoo1990h}), sampling uncertainty has been computed by treating each measurement as independent and identically distributed, yielding\footnote{Known as Bessel's correction, the $K-1$ in the denominator accounts for the bias in the estimation of sampling uncertainty.}
\begin{equation}
\text{Sampling uncertainty} = \sqrt{\frac{\sum_{i=1}^{K}\left(T_{i}-\bar{T}\right)^{2}}{K-1}},
\label{equ:sampling_uncertainty}
\end{equation}
where $K$ is the number of probes, $\bar{T}$ is the numeric average and $T_i$ represents an individual measurement. Note that \eqref{equ:sampling_uncertainty} is simply the formula for the standard deviation. In some reports, sampling uncertainty is also broken down into \emph{radial sampling error}, \emph{circumferential sampling error} and \emph{probe mislocation} (see Table 8.1-2 in \cite{pianko1983propulsion}). 

\begin{figure}
\centering
\includegraphics[width=13.0cm]{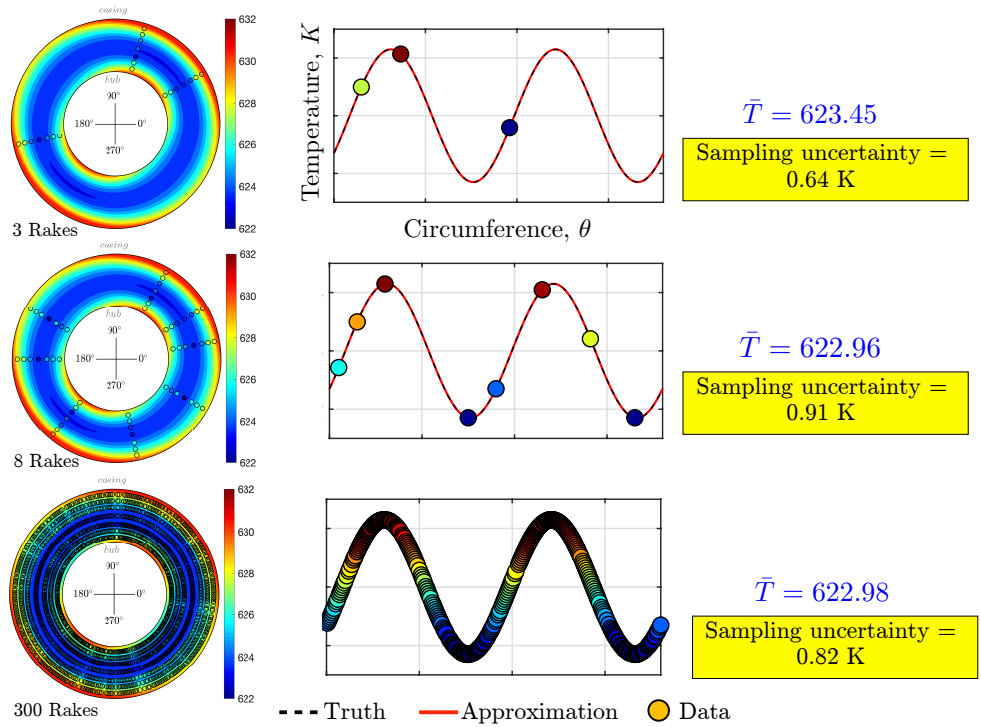}
\caption{Sampling uncertainty calculation (using \eqref{equ:sampling_uncertainty}) for an assumed engine with a single harmonic, which can be exactly captured by 3 rakes. Shown are the results for 3, 8 and 300 rakes. In none of the cases is the sampling uncertainty 0.0 K.}
\label{fig:modes}
\end{figure}

Here, we argue that such metrics (as in \eqref{equ:sampling_uncertainty}) fall short of providing a true representation of what the \emph{sampling uncertainty} is. Consider the spatial distribution in Fig.~\ref{fig:modes}, which comprises of a single circumferential harmonic of frequency 2. If we focus only on the circumferential variation at mid-span, then using least squares, three rakes should be adequate for capturing this pattern accurately\footnote{Assuming no noise, of course.} We study the \emph{sampling uncertainty} for three different cases, $K=3, 8$ and $300$. If we set $K=3$ and set $\bar{T}$ to be the numeric average of the three rakes (at mid-span), then we obtain a sampling uncertainty (standard deviation) of $\pm 0.64 K$. Note that both the average and the uncertainty is dependent on the circumferential location of the rakes; probes that are located at the peaks and troughs will likely yield different estimates compared to probes that are placed elsewhere. 

If we increase the number of rakes from 3 to 8, even though the spatial pattern is still exactly captured, the sampling uncertainty increases to $\pm 0.91$. Our expectation however, is that for both these cases the sampling uncertainty should be zero. In fact, if we further increase the number of rakes to 300, the sampling uncertainty is still not zero. Thus, $\bar{T}$ cannot be a fixed value, but must vary depending on our spatial location, because the underlying spatial pattern itself is not uniform. 

Thus, sampling uncertainty as a quantitative metric cannot be expressed in the absence of some spatial representation of its associated scalar field. Furthermore, any data-driven approximation of the spatial field will undoubtedly depend on the precision of the individual measurements themselves. These notions warrant further study into the definition of sampling and measurement imprecision. 

The remainder of this paper is structured as follows. In sec.~\ref{sec:lit} we establish why engine temperature uncertainty quantification is of paramount importance and in sec.~\ref{sec:prior} survey prior attempts aimed at recognizing and quantifying sampling and measurement uncertainties. This is followed by a detailed uncertainty quantification study of our model in sec.~\ref{sec:uq} using ideas from classical statistics. It is in this section that we offer new definitions of \emph{spatial sampling uncertainty} and the \emph{measurement imprecision uncertainty}. Numerical results based on previously introduced engine extracts (see Table 1 in part I) then follow in secs.~\ref{sec:resultsA} and \ref{sec:resultsB}.

\section{Thermodynamic uncertainty quantification}
\label{sec:lit}
In this section we briefly study the impact of an uncertainty in the averaged temperature values. Our aim is two-fold. First, to quantitatively demonstrate the impact of an uncertainty in the averaged temperature value\footnote{Here we do not distinguish between mass and area average. Moreover, our approach is invariant to whether one uses temperature or pressure values.} thereby motivating the remainder of our paper. Second, to illustrate the use of the \emph{law of propagation of uncertainty} and \emph{Monte Carlo}. 

\subsection{Taylor series expansion}
Consider the textbook definition of isentropic efficiency $\eta = \eta \left( \boldsymbol{z} \right) $, where the parameters 
\begin{equation}
\boldsymbol{z} = \left(T_{01}, T_{02}, P_{01}, P_{02}, \gamma \right).
\end{equation}
Here $\gamma$ is the specific heat capacity ratio and $T$ and $P$ represent the stagnation temperatures and pressures at the stations corresponding to the subscript in the terms, i.e.~$01$ for the inlet and $02$ for the exit. For a turbine this is expressed as 
\begin{equation}
\eta\left(\boldsymbol{z}\right)=\frac{T_{01}-T_{02}}{T_{01}\left(1-\left(\frac{P_{02}}{P_{01}}\right)^{\frac{\gamma-1}{\gamma}}\right)}.
\end{equation}
Let us assume that mean values of $\boldsymbol{z}$ are known and given by $\boldsymbol{\bar{z}} = \left( \bar{T}_{01}, \bar{T}_{02}, \bar{P}_{01}, \bar{P}_{02}, \bar{\gamma} \right)$. Then, the first-order Taylor series expansion of efficiency---following the uncertainty propagation workflow in Coleman and Steele \cite{coleman1995engineering}---about these mean values is given by
\begin{equation}
\eta\left(\boldsymbol{z}\right)\approx\eta\left(\bar{ \boldsymbol{z}}\right)+\sum_{i=1}^{5}\frac{\partial\eta}{\partial z_{i}}\left(z_{i}-\bar{z}_{i}\right),
\label{eff}
\end{equation}
where $z_i$ denotes the $i^{\text{th}}$ parameter in $\boldsymbol{z}$. Taking the expectation on both sides of \eqref{eff} yields
\begin{align}
\begin{split}
\bar{\eta} \left( \boldsymbol{z} \right) = \mathbb{E}\left[\eta\left(   \boldsymbol{z}  \right)\right] & \approx   \mathbb{E}\left[\eta\left(\boldsymbol{\bar{z}} \right) \right]  +   \sum_{i=1}^{5}\mathbb{E}\left[\frac{\partial\eta}{\partial z_{i}}\left(z_{i}-\bar{z}_{i}\right)\right] \\
&=  \eta\left(\boldsymbol{\bar{z}} \right)  + \sum_{i=1}^{5}\frac{\partial\eta}{\partial z_{i}} \underbrace{  \mathbb{E}\left[\left(z_{i}-\bar{z}_{i}\right)\right]}_{=0} \\
& = \eta \left( \boldsymbol{\bar{z}} \right).
\label{equ_eff_uq}
\end{split}
\end{align}
Unsurprisingly, this implies that the expectation in efficiency can be approximated by plugging the values in $\bar{\boldsymbol{z}}$ into \eqref{eff}. Next, we compute the variance in $\eta\left(\boldsymbol{z} \right)$,
\begin{align}
\begin{split}
\sigma^{2}\left[\eta\left( \boldsymbol{z} \right)\right] & \approx\mathbb{E}\left[\left(\eta\left( \boldsymbol{z} \right)-\eta\left(\bar{ \boldsymbol{z} }\right)\right)^{2}\right] \\
& = \mathbb{E}\left[\left(\sum_{i=1}^{5}\frac{\partial\eta}{\partial z_{i}}\left(z_{i}-\bar{z}_{i}\right)\right)^{2}\right] \\
& = \mathbb{E}\left[\sum_{i=1}^{5}\left(\frac{\partial\eta}{\partial z_{i}}\right)^{2}\left(z_{i}-\bar{z}_{i}\right)^{2}\right. \\
& \left.+2  \sum_{i=1}^{4} \sum_{j=i+1}^{5}   \left(\frac{\partial\eta}{\partial z_{i}}\right)\left(\frac{\partial\eta}{\partial z_{j}}\right)\left(z_{i}-\bar{z}_{i}\right)\left(z_{i}-\bar{z}_{j}\right)\right] \\
& = 2 \sum_{i=1}^{5} \sum_{j=1}^{5}  \left(\frac{\partial\eta}{\partial z_{i}}\right)\left(\frac{\partial\eta}{\partial z_{j}}\right) \text{cov}\left(z_{i},z_{j}\right),
\end{split}
\label{equ_variance_taylor}
\end{align}
where $\text{cov}$ denotes the covariance. Should precise information on the correlation between $z_i$ and $z_j$ be unknown, one can construct a correlation matrix of values of $z_i$ and $z_j$ from sensor data to inform $\text{cov}\left(z_{i},z_{j}\right)$; this is the approach used in Monte Carlo sampling.

\subsection{Monte Carlo sampling}
Generating independent identically distributed (iid) samples for a Monte Carlo study is a capability that is available in most commercial statistics codes and open-source libraries. For multiple uncertainties, these random samples are prescribed to have a certain mean and covariance matrix while the individual \emph{marginal distributions} have a certain probability density function. Techniques for generating multivariate correlated random samples with non-Gaussian marginals typically require the use of transforms---e.g., Nataf transform and Rosenblatt transform (see page 109 in \cite{smith2013uncertainty})---or more generally the use of copulas (see \cite{lebrun2009rosenblatt, trivedi2007copula}). Alternatively, one can use acceptance-rejection based algorithms (see page 342 in \cite{casella2004generalized}), where one generates samples from a \emph{proposal distribution} iteratively based on whether they fall within a \emph{target distribution}. Both these distributions can be set based on the marginal distributions and the correlations required. 

\begin{table}[h]
\begin{center}
\caption{Individual parameter uncertainties. }
\begin{tabular}{|c|c|}
  \hline		
 Parameter & Uncertainty,  $\pm \sigma \left(p_i \right)$ \\ \hline
 \hline
$T_{01}$ &  2.4 K  \\
$T_{02}$  &  1.4 K \\
$P_{01}$ &  600 Pa\\
$P_{02}$ & 100  Pa\\
$\gamma$ & 0.001 \\
  \hline  
\end{tabular}
\label{table:individual_uncertainties}
\end{center}
\end{table}

\subsection{Uncertainty contributions and correlations}
\begin{figure}
\centering
\includegraphics[width=9.0cm]{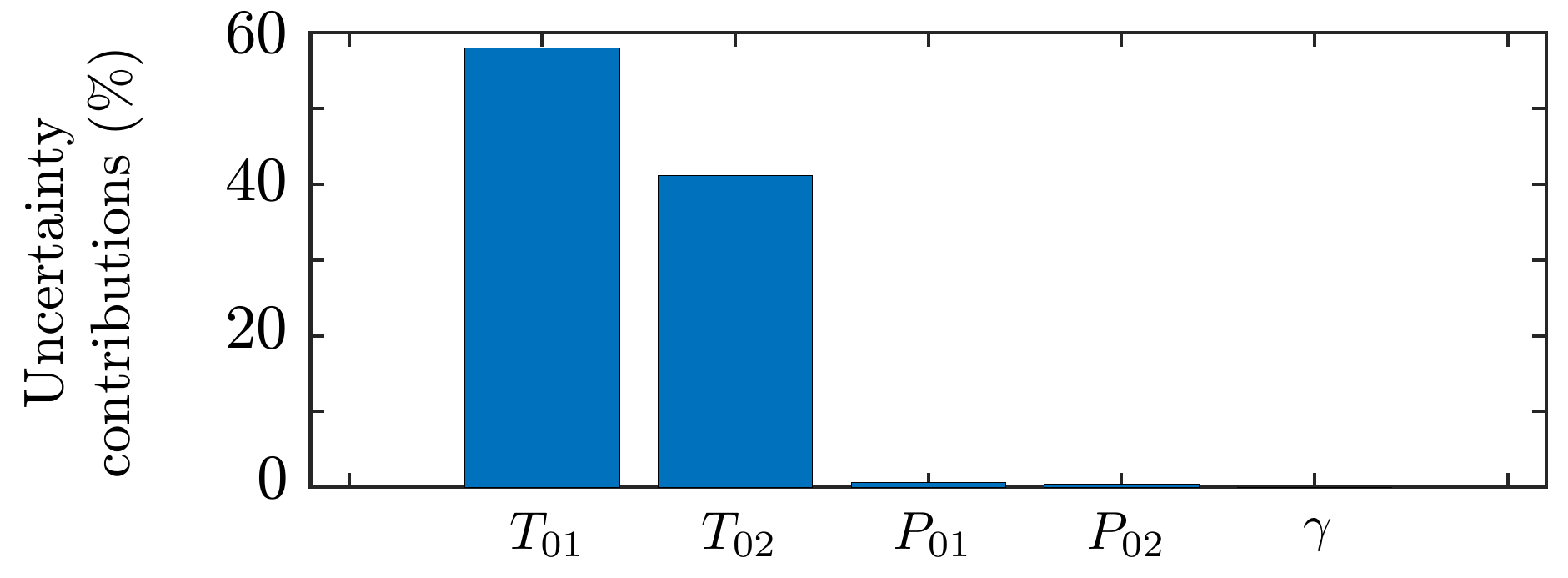}
\caption{Uncertainty contributions of pressures, temperatures and specific heat capacity ratio to efficiency.}
\label{fig:uncertainty_contributions}
\end{figure}

We now re-visit \eqref{equ_variance_taylor} with the aim of \emph{ranking} the influence of the various uncertainties. Emulating the lines of thought pursued by V\'{a}zquez and S\'{a}nchez \cite{vazquez2003temperature}, we assign values to both the partial derivative terms and the individual variances. For computing the partial derivative terms we use aero-industry engine representative values of temperatures and pressures, and for the individual variances we use the values shown in Table \ref{table:individual_uncertainties}. This permits us to compute the individual uncertainty contributions,
\begin{equation}
\left(\frac{\partial\eta}{\partial p_{i}}\right)^{2}  \sigma^{2}  \left(  p_{i}  \right).
\end{equation}
Correlations between the different flow quantities are initially set to zero, implying that they are independent of each other. Next, we divide these individual \emph{uncertainty contributions} by the total uncertainty in efficiency, which we compute to be $\pm 1.2\%$ ($2 \sigma$), and plot them in Fig.~\ref{fig:uncertainty_contributions}. The conclusion is clear: uncertainties in temperatures are sizeably more important than uncertainties in pressures and the specific heat capacity ratio for the turbine example case considered.

Nearly identical results can be obtained when using Monte Carlo sampling for this \emph{uncertainty propagation}. We generate an ensemble of multivariate Gaussian distributions with means corresponding to the aero-industry representative values and standard deviations as shown in Table~\ref{table:individual_uncertainties}, with 500,000 independent samples\footnote{A convergence study was carried out to ensure that the number of samples was sufficient.}. For each of these samples, where each sample is a vector of 5 values, we use \eqref{eff} to compute the isentropic efficiency. 

\begin{figure}
\centering
\includegraphics[width=8cm]{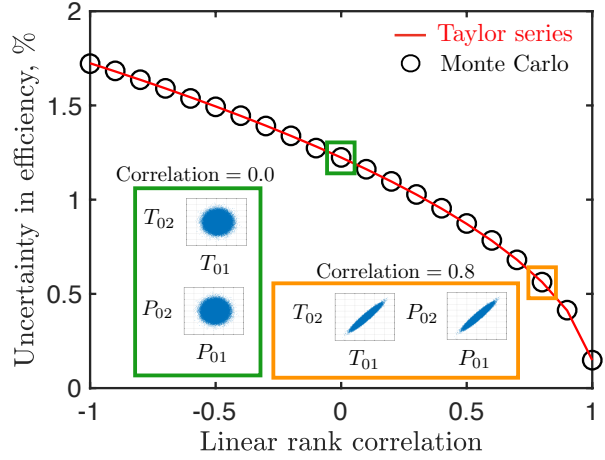}
\caption{Uncertainty in efficiency as a function of correlations in pressures and temperatures with correlations.}
\label{fig:taylor_series}
\end{figure}

We repeat the uncertainty propagation exercise, using both the Taylor series expansion and Monte Carlo, but now enforcing correlations among the inputs. The two stagnation temperatures are assumed to be correlated with each other for this study, as are the two pressures. In Fig.~\ref{fig:taylor_series} we plot the resulting uncertainty in efficiency for a range of different Pearson rank correlation values. These scalar values are bounded between -1 and 1 and, for two variables $z_1$ and $z_2$, defined by
\begin{equation}
\rho_{z_1, z_2} = \frac{\text{cov} \left( z_1, z_2    \right)    }{ \sigma_{z_1} \sigma_{z_2} }.
\end{equation}
Thus, for each of the circular markers in Fig.~\ref{fig:taylor_series}, we generate samples that have a Pearson rank correlation matrix of the form:
\begin{equation}
\boldsymbol{\rho} = \begin{blockarray}{cccccc}
T_{01} & T_{02} & P_{01} & P_{02} & \gamma \\
\begin{block}{(ccccc)c}
  1 & \rho_{T_{01}, T_{02}} & 0 & 0 & 0 & T_{01} \\
 \rho_{T_{01}, T_{02}} & 1 & 0 & 0 & 0 & T_{02} \\
  0 & 0 & 1 & \rho_{P_{01}, P_{02}} & 0 & P_{01} \\
  0 & 0 & \rho_{P_{01}, P_{02}} & 1 & 0 & P_{02} \\
  0 & 0 & 0 & 0 & 1 & \gamma \\
\end{block}
\end{blockarray}
\end{equation}
In other words, for a linear rank correlation value of 0.8 (see the horizontal axis in Fig.~\ref{fig:taylor_series}), we set
\begin{equation}
\rho_{T_{01}, T_{02}} = \rho_{P_{01}, P_{02}} = 0.8.
\end{equation}

Our conclusions follow those of V\'{a}zquez and S\'{a}nchez \cite{vazquez2003temperature} who show that positive correlations among the input parameters (particularly the temperatures) have the effect of reducing the overall uncertainty in efficiency. On the other hand, if the flow quantities are negatively correlated with each other, an increase in efficiency uncertainty is expected. In fact, in practice, as the thermocouples and temperature measurement systems are calibrated with respect to the same standard, $T_{01}$ and $T_{02}$ will correlate \cite{vazquez2003temperature}.

There are two key take-home messages here. The first is that correlations are extremely important and that if we seek to reduce the uncertainty in efficiency measurement, then we must rigorously account for both the individual uncertainty contributions and their correlations. The second is that one can extract exchange rates from Figs.~\ref{fig:uncertainty_contributions} and \ref{fig:taylor_series} to quantify the repercussions of an uncertainty in a flow quantity. In fact, such exchange rates are often used in more comprehensive, engine-wide, preliminary design and analysis tools. 

\section{Sampling and measurement imprecision in literature}
\label{sec:prior}
Before we detail our new data-centric framework, it will be worthwhile to examine prior work focused on metrics for sampling and measurement imprecision. 

\subsection{Sampling uncertainty}
Early efforts \cite{abernathy1973handbook} aimed at calculating uncertainty budgets in engines grouped uncertainties associated with spatial sampling into ``other effects'' along with sensor system errors, errors of method, and non-instrumentation errors. Vleghert et al.  \cite{vleghert1989} recognized that such broad categorizations came with limitations, but their work still adopted the widely used formulations in \cite{abernathy1973handbook}. More recently, Bonham et al. \cite{bonham2017combination} re-visited the issue of spatial sampling, highlighting the fact that far fewer measurements will be required towards the inlet of a compressor, owing to the uniformity of the flow. However, towards the rear of the compressor, wakes from the upstream components will cause spatial distortions in the flow-field that will alter time-average measurements. Further, the authors assume a harmonic profile with a period of a half in the radial direction and state that at least 7 measurements in the radial direction are required to guarantee a compressor efficiency uncertainty to be less than $\pm 0.1 \%$. They then state that a similar analysis shows that 5 measurements are required in the circumferential direction (see page 4 in \cite{bonham2017combination}) for their specific experiment.

Other sampling uncertainty calculations in literature include example 10.4 of the ASME PTC 19.1 test codes \cite{dieck2005test}, where the uncertainty in the efficiency of a compressor is calculated. \emph{Spatial averaging} (see Table 10.4.3.2-1 and Table 10.4.3.2-1 in \cite{dieck2005test}) is listed as the largest uncertainty contributor to the exit stagnation pressure and is the second-largest uncertainty contributor to the exit stagnation temperature. The technique for computing this spatial average uncertainty is \eqref{equ:sampling_uncertainty}, which once again seems misleading\footnote{We note the comment made by Dieck (see page 188 in \cite{dieck2007measurement}), where in reference to spatial sampling uncertainty the author states that ``One way to evaluate the effects of space [...] is to see your local statistician...''. Our interpretation of this statement is the inherent difficulty in quantifying spatial sampling uncertainty. This motivated writing this paper with our second co-author.}. 

\subsection{Measurement imprecision}
The AGARD-AR-245 report \cite{saravanmuttoo1990h} decomposes temperature uncertainties into factors based on the design of the probe---e.g., velocity error, conduction error, convection error, radiation error (see page 85 in \cite{saravanmuttoo1990h})---static calibration uncertainties, data acquisition uncertainties, data reduction (curve-fit) uncertainties and sampling uncertainties. We refer to the totality of these uncertainties, with the exception of the last one, as measurement uncertainties.

The velocity error, which we hereby refer to as the \emph{dynamic calibration uncertainty}, arises in a temperature probe as the air is not brought to rest adiabatically. In other words, the measured stagnation temperature is below the true stagnation temperature of the fluid. This difference is accounted for by an empirical recovery factor, the value of which typically depends on the Mach number of the incoming flow (see Moffat \cite{moffat1962gas}). Next, we have the static calibration uncertainty, which accounts for uncertainties in the calibration of the probe, often determined by the precision of the individual thermocouples (or platinum resistance thermometers) themselves. 

The overall uncertainty calculations in AGARD-AR-245 (see page 137 in \cite{saravanmuttoo1990h}) are computed using the root sum square method on the \emph{test vehicle uncertainty}, \emph{pressure and temperature measurement uncertainty} and \emph{sampling uncertainty}. In other words, sources of measurement uncertainty are deemed to not have an impact on the spatial flow distortion pattern; this underscores the need for further study.
\section{Propagating uncertainties}
\label{sec:uq}
In this section we layout our framework for propagating uncertainties in the least squares based model presented in part I (see \cite{seshadri2019a}). 
\subsection{Theoretical framework}
We will be borrowing some of our notation from \cite{seshadri2019a}. Let $N$ be the number of rakes and $M$ be the number probes per rake. The mean of each of the $N \times M$ measurements is in $\boldsymbol{\mu}_{\mB} \in \mathbb{R}^{N \times M}$ and the associated covariances of the measurements are incorporated in $\boldsymbol{\varSigma}_{\mB} \in \mathbb{R}^{NM \times NM}$, i.e.~$\mB \sim \mathcal{N} \left( \boldsymbol{\mu}_{\mB} , \boldsymbol{\varSigma}_{\mB} \right)$, where the symbol $\mathcal{N} \left( \cdot, \cdot \right)$ indicates a multivariate Gaussian distribution with the mean as the first argument and the covariance as the second. We assume that sufficient information on the precision of each probe, recovery factors and the standard against which the calibrations are applied is known to justify values for $\boldsymbol{\mu}_{\mB}$ and $\boldsymbol{\varSigma}_{\mB}$. The Fourier matrix (see (2) in \cite{seshadri2019a}) is given by $\mA \in \mathbb{R}^{N \times \left(2k + 1 \right) }$ where $k$ is the number of harmonics. In what follows, we aim to propagate this uncertainty through the model to infer what the consequent uncertainty in the spatial temperature field is. Defining the \emph{pseudoinverse} of $\mA$ as
\begin{equation}
\mP = \left( \mA^{T} \mA  \right)^{-1} \mA^{T}  \; \; \; \text{where} \; \; \; \mP \in \mathbb{R}^{\left(2k+1 \right) \times N},
\end{equation}
and the mean and the covariance in the coefficients $\mX \in \mathbb{R}^{\left(2k+1\right) \times M}$---a non-deterministic analogue to the \emph{multivariate regression problem} (see (3) in part I)---is given by
\begin{equation}
\boldsymbol{\mu}_{\mX}= \mP \boldsymbol{\mu}_{\mB} \; \; \; \; \text{and} \; \; \; \; \boldsymbol{\varSigma}_{\mX}=  \left( \mI_{M} \otimes \mP  \right) \boldsymbol{\varSigma}_{\mB} \left( \mI_{M} \otimes \mP  \right)^{T},
\end{equation}
where $\boldsymbol{\mu}_{\mX}$ has the same dimensions as $\mX$ and $\boldsymbol{\varSigma}_{\mX} \in \mathbb{R}^{\left(2k+1 \right)M \times \left(2k+1 \right)M}$. Note that here $\mI_{M} \in \mathbb{R}^{M \times M}$ is the identity matrix. The mean and the covariance in the output use these expressions; these are given by
\begin{equation}
\boldsymbol{\mu}_{\mF}=  \mA \boldsymbol{\mu}_{\mX} \; \; \; \; \text{and} \; \; \; \; \boldsymbol{\varSigma}_{\mF}=  \left(\mI_{M} \otimes \mA  \right) \boldsymbol{\varSigma}_{\mX} \left(\mI_{M} \otimes \mA  \right)^{T},
\label{equ_f}
\end{equation}
where $\boldsymbol{\mu}_{\mF}$ has the same dimensions as $\boldsymbol{\mu}_{\mB}$ and $\boldsymbol{\varSigma}_{\mF}$ has the same dimensions as $\boldsymbol{\varSigma}_{\mB}$; note that in \eqref{equ_f} . To estimate the predicted mean and predicted variance at any point over the domain, we use Equation 9 from \cite{seshadri2019a}, which is
\begin{equation}
T \left(r, \theta \right) = \vv^{T}\left(r\right) \mU  \mX^{T} \va \left( \theta \right).
\end{equation}
The predictive mean at any point conditioned upon the measurements is simply given by replacing $\mX$ with $\boldsymbol{\mu}_{\mX}$, yielding 
\begin{equation}
\mathbb{E} \left[T | \boldsymbol{\mu}_{\mB} \right] = \vv^{T}\left(r\right) \mU  \boldsymbol{\mu}_{\mX}^{T} \va \left(\theta\right).
\end{equation}
Average values of the predictive mean over the entire spatial domain can be readily obtained by integrating these quantities in polar coordinates, as will be shown in the next subsection.

From \eqref{equ_f}, the mean and covariance in the residuals $\left( \boldsymbol{\mu}_{\mR}, \boldsymbol{\varSigma}_{\mR} \right)$ can be written as
\begin{equation}
\boldsymbol{\mu}_{\mR}=  \boldsymbol{\mu}_{\mF} - \boldsymbol{\mu}_{\mB} \; \; \; \; \text{and} \; \; \; \; 
\label{predictive_mean}
\end{equation}
\begin{equation}
\boldsymbol{\varSigma}_{\mR}=  \left( \mI_{M} \otimes \left( \mA \mP - \mI_{N}  \right)  \right) \boldsymbol{\varSigma}_{\mB}  \left( \mI_{M} \otimes \left( \mA \mP - \mI_{N}  \right)  \right)^{T}.
\label{predictive_var}
\end{equation}
It is important to note that in standard linear least squares models (see page 47 in \cite{friedman2009elements} or page 15 in \cite{faraway2016linear}) one assumes that the errors are iid, and characterized by a $\mathcal{N} \left(0, \sigma^2 \mI \right)$ distribution. In \emph{generalized least squares} models (see page 89 in \cite{faraway2016linear}) the iid assumption is relaxed, with an error of the form $\mathcal{N} \left(0, \boldsymbol{\varSigma} \right)$. The residuals above fit neither of these forms and thus we require further theory to ascertain their moments. 

While the distributions associated with $\mX$, $\mF$ and $\mR$ are multivariate Gaussian, the distribution associated with the norm of the residual is not Gaussian. Our goal is to obtain an analytical expression for the distribution of the quantity
\begin{equation}
\epsilon_{p}^{2} = \frac{1}{NM} \left\Vert \mA\mX-\mB\right\Vert _{2}^{2},
\end{equation}
given the uncertainty in $\mB$. This expression can be recast as a quadratic form associated with two Gaussians, which admits a non-central chi-square distribution under certain criteria (see Theorem 7.3 from Rao and Mitra \cite{rao1972generalized} for details). The criterion relevant to us is the assumption that 
\begin{equation}
\boldsymbol{\varSigma}_{\mB} = \sigma^2_{b} \mI.
\label{equ_raocriterion}
\end{equation} 
This results in the non-central chi-square distribution $\chi^2 \left(g, \phi \right)$ where $g$ represents the degrees of freedom and $\phi$ the non-centrality parameter; these parameters are computed via
\begin{equation}
g = \text{rank} \left( \boldsymbol{\varSigma}_{\mR} \right)\; \; \; \; \text{and}\; \; \; \; \phi = \text{vec}\left(\boldsymbol{\mu}_{\mR} \right)^{T} \boldsymbol{\varSigma}_{\mR}^{-} \text{vec}\left(\boldsymbol{\mu}_{\mR} \right),
\label{equ:chisquare_params}
\end{equation}
where $\boldsymbol{\varSigma}_{\mR}^{-}$ is the pseudoinverse of $\boldsymbol{\varSigma}_{\mR}$. The symbol $\text{vec}\left( \cdot \right)$ denotes a  column-wise vectorized version of the argument. The moments associated with the error are then given by
\begin{equation}
\mu\left(\varepsilon_{p}^{2}\right)=\frac{\sigma_{b}^{2}}{NM}\left(g+\phi\right) \; \; \; \text{and} \; \; \; \sigma^{2}\left(\varepsilon_{p}^{2}\right)=\frac{\sigma_{b}^{2}}{NM}\left(2g+4\phi\right).
\label{equ:uncertainty_in_errors}
\end{equation}
In closing this theoretical discussion, we summarize our main points below:
\begin{enumerate}
\item If the uncertainty in the thermodynamic measurements are Gaussian and uncorrelated, then \eqref{equ:uncertainty_in_errors} can be used to determine the mean and variance in $\epsilon_{p}^2$. 
\item Furthermore, in this case one can obtain the probability density of the quantity
\begin{equation}
\frac{NM}{\sigma_{b}^2} \epsilon^2_{p},
\label{equ_error_pdf}
\end{equation}
(which will be a non-central chi-square distribution), but not the probability density of $\epsilon^2_{p}$ alone. 
\item If the uncertainties in the temperatures are Gaussian and correlated, or from any other distribution, then one should carry out a Monte Carlo study to ascertain the moments in $\epsilon_{p}^2$. 
\item If the uncertainties in the temperatures are Gaussian and correlated, one can still utilize \eqref{equ_f} to obtain the predictive mean and predictive variance values.
\end{enumerate}

\subsection{Area average computations}
The mean and variance in the area average value of temperature (or pressure) can be obtained by integrating the spatial mean and spatial covariance fields. Thus, the expected area average value for the case considered in part I is given by
\begin{align}
\begin{split}
\mathbb{E} \left[ T_{avge}  \right] & = \frac{1}{\pi\left(r_{o}^{2}-r_{i}^{2}\right)}\int_{r_{i}}^{r_{o}}  \int_{0}^{2 \pi} \vv^{T} \left(r\right) \mU \boldsymbol{\mu}_{\mX}^{T} \va \left(\theta\right)   r dr d\theta \\
& =  \frac{2}{\left(r_{o}^{2}-r_{i}^{2}\right)}\int_{r_{i}}^{r_{o}}  r \vv^{T} \left(r\right) \mU\left[\begin{array}{cc}
\boldsymbol{\mu}_{1}^{T} & \boldsymbol{0}
\end{array}\right] dr
\end{split}
\label{equ:uq_area_new}
\end{align}
where, as before (see section 4.3 in \cite{seshadri2019a}), we have used the fact that the harmonic terms in $\va$  become zero when integrated from $0$ to $2\pi$. Here $\boldsymbol{\mu}_{1}^{T}$ corresponds to the first column of $\boldsymbol{\mu}_{\mX}^{T}$. To estimate the area average covariance between any two sets of points $\vz = \left( \theta, r \right)$ and $\vz' = \left( \theta', r' \right)$, we write
\begin{align}
\begin{split}
\text{Cov}_{\left( \vz, \vz' \right) } \left[ T_{avge} \right] & = \int_{r_{i}}^{r_{o}}  \int_{0}^{2 \pi}  \int_{r_{i}}^{r_{o}}  \int_{0}^{2 \pi}  \left(   \vv\left(r \right)^{T} \mU \left( \mI_{M} \otimes \va\left( \theta \right) \right)^{T} \boldsymbol{\varSigma}_{\mX} \left( \mI_{M} \otimes \va\left( \theta' \right) \right) \mU^{T} \vv\left(r' \right)   \right)   \\
& \cdot  \frac{r' dr' d\theta' r dr d\theta}{\pi^2\left(r_{o}^{2}-r_{i}^{2}\right)^2 } \\
& = \frac{4}{\left(r_{o}^{2}-r_{i}^{2}\right)^2} \int_{r_{i}}^{r_{o}}  \int_{r_{i}}^{r_{o}}  \left(   \vv\left(r \right)^{T} \mU \left( \mI_{M} \otimes \vn \right)^{T} \boldsymbol{\varSigma}_{\mX} \left( \mI_{M} \otimes \vn \right) \mU^{T} \vv\left(r' \right)   \right)   r r' dr dr',
\end{split}
\label{equ:uq_area_var}
\end{align}
where $\vn \in \mathbb{R}^{2k + 1}$ is a vector of zeros, except for the first entry, which has a value of one. To compute the variance, we simply set $r=r'$. It should be clear from \eqref{equ:uq_area_var} that the variance in the area average temperature value will not be a function of the harmonic terms in $\va \left( \theta \right)$, but only of the constant term, i.e., the first one.

\subsection{Metrics for sampling and measurement imprecision}
So what metrics can we use to characterize spatial sampling and measurement imprecision uncertainty? Based on the formulations above, we propose the following two metrics.
\begin{enumerate}
\item \underline{Spatial sampling uncertainty}: Assuming one can approximate the key Fourier modes (using techniques detailed in part I), the sampling uncertainty is given by:
\begin{equation}
\epsilon_{p}^2 = \frac{1}{NM - 1} \left\Vert \mA\mX- \boldsymbol{\mu}_{\mB} \right\Vert _{2}^{2},
\end{equation}
where, as before, the matrix $\mA$ encodes assumptions on which harmonics to use. Large values of $\epsilon_{p}^2$ will be indicative of a deficient choice (and number) of the harmonic frequencies. 
\item \underline{Measurement imprecision uncertainty}: We define measurement imprecision uncertainty to be the contribution of individual measurement uncertainties on the spatial flow pattern, given by
\begin{equation}
\epsilon_{m}^2 = \mu \left( \epsilon_{p}^2 \right) - \epsilon_{p}^2,
\end{equation}
where $\epsilon_{p}^2$ is the sampling uncertainty and $\mu \left( \epsilon_{p}^2 \right)$ (see \eqref{equ:uncertainty_in_errors}) is the total uncertainty---combining both the effect of assumed spatial harmonics and measurement imprecision. It should be noted that 
 \begin{equation}
\underset{\sigma_{b}\rightarrow0}{ \text{limit} } \; \; \mu\left(\epsilon_{p}^{2}\right)=\epsilon_{p}^{2},
\end{equation}
implying that, if there is no uncertainty in the measurements, then $\epsilon_{m}^2$ will be zero.
\end{enumerate}
Both the uncertainty metrics stated above are given as variances in the units of the quantity they measure. We emphasize that one shortcoming in our analytical derivations is that we cannot offer an analytical form for $\mu \left( \epsilon_{p}^2 \right)$ when the measurement uncertainties are correlated. In this case Monte Carlo sampling must be used.
\section{Numerical results}
\label{sec:resultsA}
In what follows we apply the formulations above on the engine extracts previously introduced in part I.
\subsection{Results on an engine extract}
In this subsection we focus on an isolated extract from Engine A and will assume that the harmonic pair $\boldsymbol{\omega} = \left(1, 4 \right)$ affords an accurate spatial representation of the temperature data. As we only have 6 rakes, we do not discern between \emph{testing} and \emph{training} data (see sec.~3.5 in part I). Our first task is to inject uncertainty into individual temperature measurements. We will assume that we are provided with the temperature uncertainties acting on each thermocouple---effectively aggregating uncertainties arising from static calibrations, dynamic calibrations and the data acquisition system. 

\begin{figure}
\centering
\includegraphics[width=11cm]{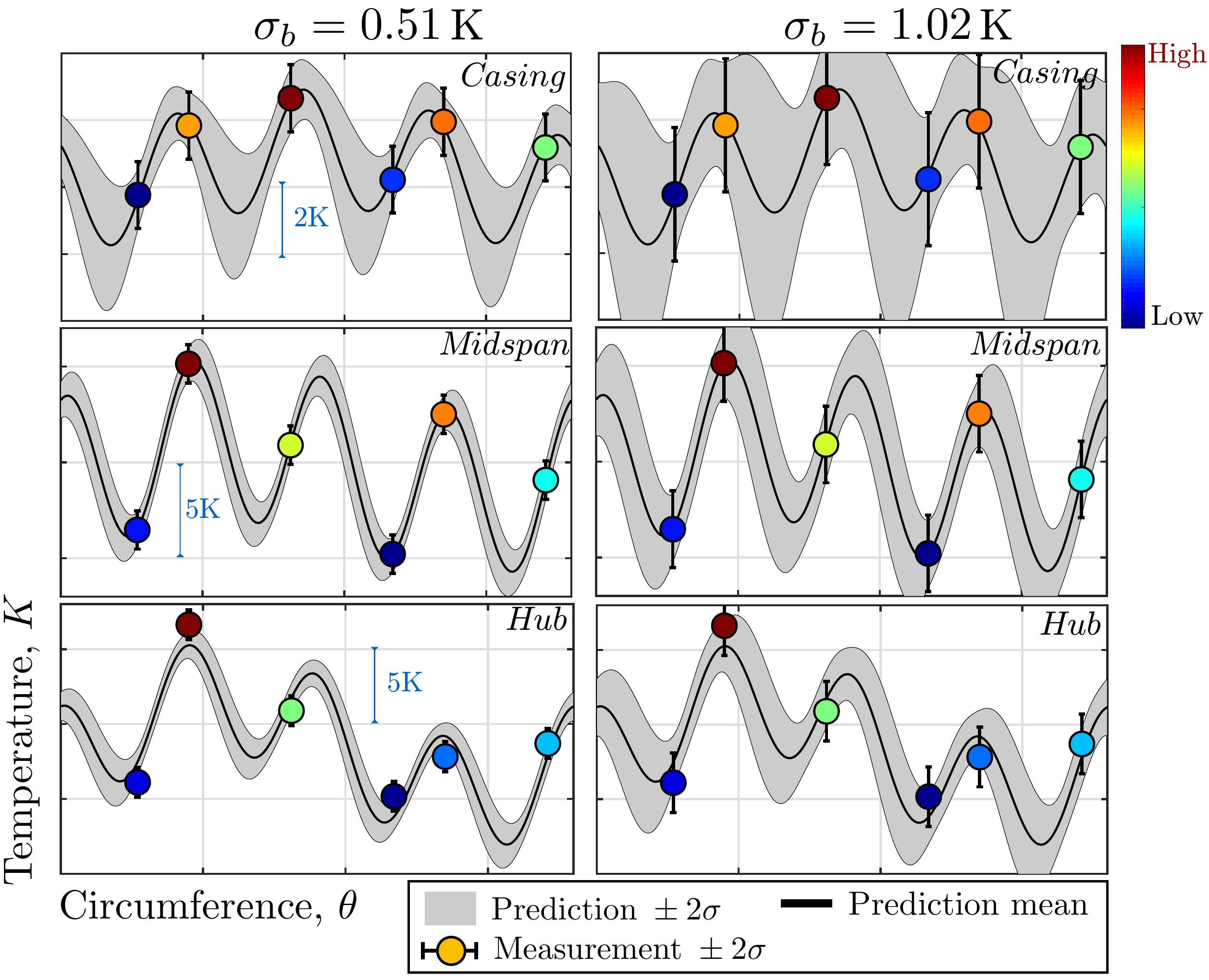}
\caption{Predictive mean and $2\sigma$ contours for uncorrelated temperature measurements with $\sigma_{b}=0.51$K (left) and $\sigma_{b}=1.02$K (right). The error bars on the measurements indicate $2 \sigma_{b}$ intervals.}
\label{fig:uq1}
\end{figure}

\begin{figure}
\centering
\includegraphics[width=11cm]{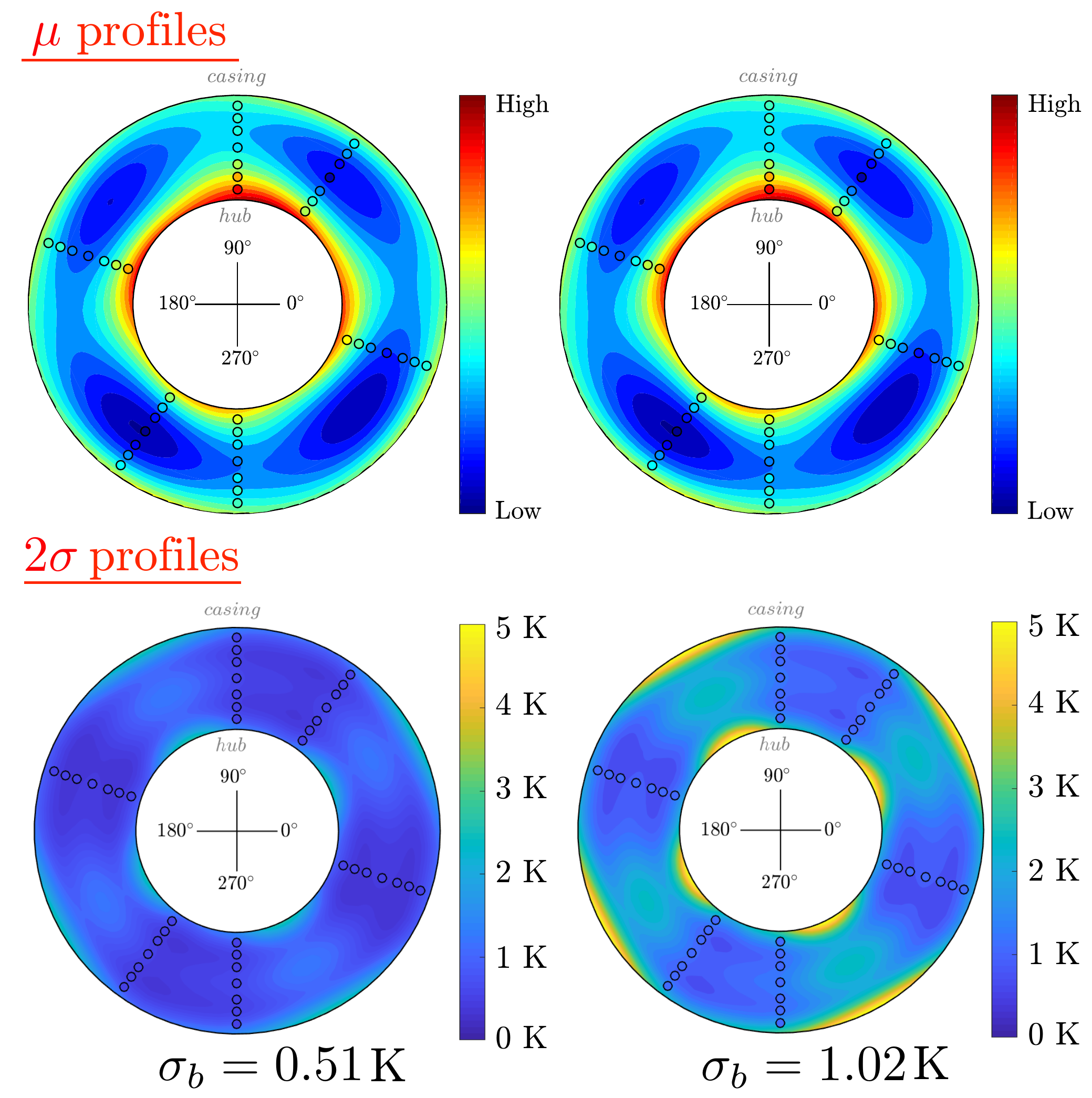}
\caption{Full spatial representations of temperature contours for uncorrelated temperature measurements with $\sigma_{b}=0.51$K (left) and $\sigma_{b}=1.02$K (right).}
\label{fig:uq2}
\end{figure}

Recall in \ref{sec:uq}, we made the assumption that our measurements had a multivariate distribution with mean $\boldsymbol{\mu}_{\mB}$ and covariance $\boldsymbol{\Sigma}$. For simplicity, we initially consider the uncorrelated case, i.e., where 
\begin{equation}
\boldsymbol{\Sigma} =\left[\begin{array}{ccc}
\sigma_{b} &  & 0\\
 & \ddots\\
0 &  & \sigma_{b}
\end{array}\right]=\sigma_{b} \mI,
\end{equation}
and thus the \emph{correlation matrix} is the identity $\mI$. To clarify, $\sigma_{b}$ is the standard deviation in the uncertainty in each probe's measurement. In Fig.~\ref{fig:uq1} we plot the predictive mean and variance contours with $\sigma_{b} = 0.51$K on the left and $\sigma_{b}=1.02$K on the right. These values result in a $2\sigma$ (95$\%$ percentile) uncertainty in individual temperature values of $\pm 1$K and $\pm 2$K respectively\footnote{For a Gaussian distribution, $95\%$ percent of the area under the curve lies within 1.96 standard deviations from the mean. Thus, we say that $2\sigma$ values are given by $\sigma_{b}$ values multiplied by 1.96}. The results shown in Fig.~\ref{fig:uq1} are the circumferential temperature variations for non-dimensional spanwise locations corresponding to 10$\%$, $50\%$ and $90\%$ which we label hub, midspan and casing respectively. These results are obtained using the analytical expressions in \eqref{predictive_mean} and \eqref{predictive_var}. The full spatial representations for these two cases are shown in Fig.~\ref{fig:uq2} along with the values of $\epsilon_{p}^2$ and $\sigma^2 \left( \epsilon_{p}^2 \right)$ in Table~\ref{table:study}. 

\begin{table}
\begin{center}
\caption{Uncertainty in error $\epsilon_{p}^2$ for an isolated extract in Engine A.}
\begin{tabular}{|c|c|c|c|}
  \hline		
 Case & $\mu \left( \epsilon_{p}^2 \right)$ & $\sigma^2 \left( \epsilon_{p}^2 \right)$ & Mean 2$\sigma$ temperature \\ \hline
 \hline
$\sigma_{b} = 0.51$K &  0.341 & 1.278 &  0.9K \\
$\sigma_{b} = 1.02$ K &  0.471 & 1.537 & 1.8K  \\
  \hline  
\end{tabular}
\label{table:study}
\end{center}
\end{table}

A few remarks on the above results are in order. The first is that a 1K uncertainty does not result in a uniformly distributed variation of 1K. Both the mean 2$\sigma$ temperature values (as shown in Table~\ref{table:study}) and the spatial variation in Figs.~\ref{fig:uq1} and ~\ref{fig:uq2} clearly illustrate this. From the latter, for the case where $\sigma_{b} = 1.02$K, implying a 2$\sigma$ value in the temperature of $\pm 2$K, a maximum 2$\sigma$ value in the spatial temperature of 5.54K was found. In both these cases the peak-to-peak variation in temperature was found to vary more towards the hub and the casing. 

\begin{figure}
\centering
\includegraphics[width=11.0cm]{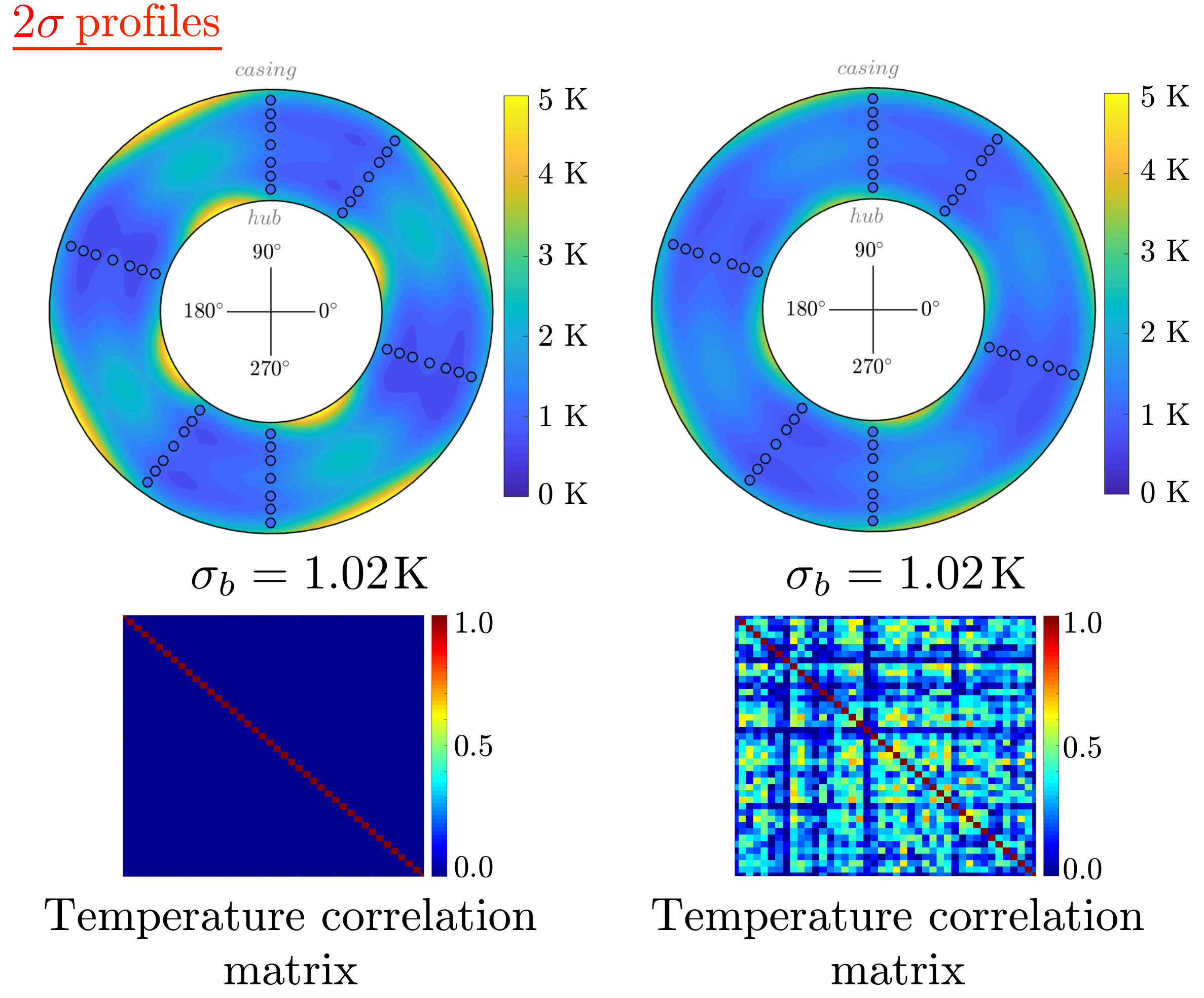}
\caption{Comparisons of $2\sigma$ profiles in temperature for $\sigma_{b}=1.02K$ with and without correlations. Note that the correlation matrix used here sets certain measurements to be independent (those that have a $\rho$ close to zero) and others to be dependent (those have a $\rho$ close to one).}
\label{fig:uq3}
\end{figure}

But what if the temperature measurement chains were calibrated such that the measurement uncertainties for each probe were correlated with the others? Fig.~\ref{fig:uq3} compares the $2\sigma$ profiles for the uncorrelated case with a correlated case for $\sigma_{b} = 1.02$K. It is clear that the peak-to-peak variation in the correlated case is reduced; here the maximum 2$\sigma$ value in the spatial temperatures is 3.9K---less than the uncorrelated value of 5.54K. Thus, the worst-case uncertainty values here can be given by the uncorrelated case. 

\subsection{Propagating uncertainties in measurements}
Recall, in section 3.5 of part I \cite{seshadri2019a}, we deployed an iterative algorithm to identify suitable harmonic pairs. Here, we repeat the same numerical exercise; however, we incorporate the uncertainty in the temperature measurements. For completeness, our \emph{non-deterministic} version of Algorithm 1 in \cite{seshadri2019a} is given below.

\begin{algorithm}
\caption{Non-deterministic brute force frequency selection for two harmonics.}
\label{alg:bruteforce2}
\begin{algorithmic}[1]
\STATE{Set $\boldsymbol{\omega}=\left(\omega_1, \omega_2 \right)$, where $\omega_1 \neq \omega_2$ and $\text{max}\left\{ \boldsymbol{\omega}  \right\} \leq 10$.}
\STATE{Obtain $\sigma_{b}$ and $\boldsymbol{\mu}_{\mB}$}
\STATE{Solve $\hat{\mX}={\text{argmin}}\left\Vert \mA\mX-\mB\right\Vert _{2}^{2}$}
\STATE{Set $\boldsymbol{\lambda} = \left(0.0001, 0.001, 0.1, 10 \right)$}
\WHILE{$\left\Vert \hat{\mX} \right\Vert _{2}\geq\beta$}
\STATE{$\lambda_i = \boldsymbol{\lambda}\left( i \right)$}
\STATE{Solve $\hat{\mX}={\text{argmin}}\left\Vert \mA\mX-\mB\right\Vert _{2}^{2} + \left\Vert \lambda_{i} \mX \right\Vert _{2}^{2}$  }
\ENDWHILE
\STATE{Compute $\boldsymbol{\mu}_{\mR}$, $\boldsymbol{\varSigma}_{\mR}$ and then $\mu \left(\varepsilon_{p}^{2} \right)$}
\RETURN $\mu \left(\varepsilon_{p}^{2} \right)$ \end{algorithmic} \end{algorithm}

\begin{figure}
\centering
\includegraphics[width=11.0cm]{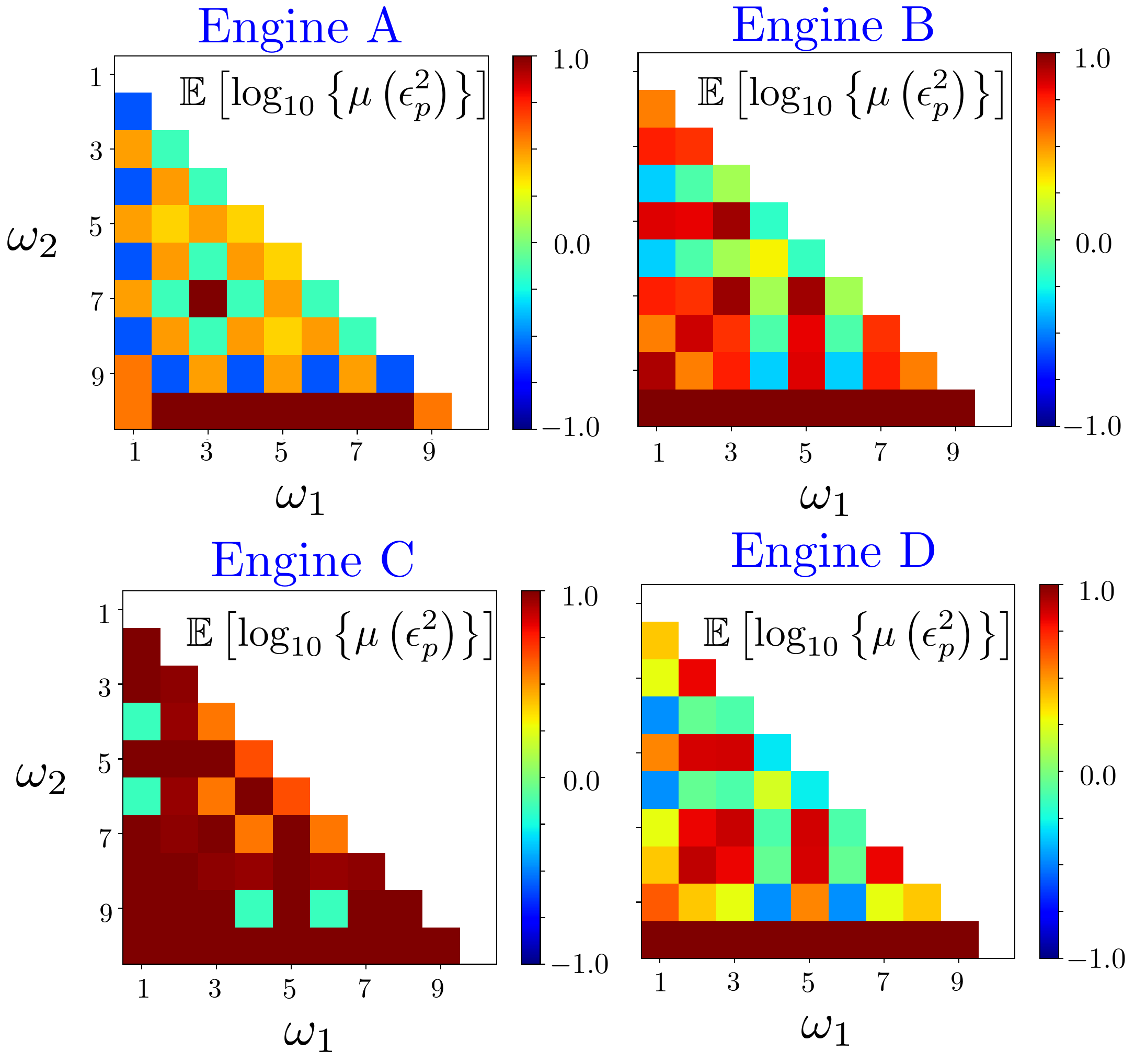}
\caption{Average values of the $\epsilon_{p}^2$ error for Engines A, B, C and D, when $\sigma_{b}=0.51$K.}
\label{fig:uq4}
\end{figure}

Once again, we emphasize that $\mu \left(\epsilon_{p}^2 \right)$ can only be computed analytically in the case where the uncertainties in individual probe measurements are Gaussian and independent. The results of Algorithm~\ref{alg:bruteforce2} are illustrated in Fig.~\ref{fig:uq4} for the independent case. The frequency pairs that yield low values of $\mu \left(\varepsilon_{p}^{2} \right)$ are the same as identified in part I, i.e., $\boldsymbol{\omega} = \left( 1, 4 \right)$, 
$\boldsymbol{\omega} = \left( 1, 6 \right)$, $\boldsymbol{\omega} = \left( 4, 9 \right)$ and $\boldsymbol{\omega} = \left( 6, 9 \right)$. 

We note that in Fig.~\ref{fig:uq4} there a few frequency pairs (such as all pairs with $\omega_2=10$) with high error values---more so than the deterministic case. This arises because in these cases $\mA$ is very poorly conditioned, and thus when we compute $\boldsymbol{\varSigma}_{\mF}$ as per \eqref{equ_f} and the subsequent terms to determine $\mu \left(\varepsilon_{p}^{2} \right)$, the ill-conditioning exacerbates. For further particulars on the importance of regularization, please see section 3.3 in the first paper.

\subsection{Propagating uncertainties in rake positions}
We now consider the problem of an uncertainty in rake positions. While rakes are typically installed at pre-determined locations, during both assembly and testing small changes in the rake position are not unexpected. Our supervised learning model permits us to propagate this uncertainty. Information on rake placement is fed through the Fourier matrix $\mA \left(\boldsymbol{\theta}, \boldsymbol{\omega} \right)$.

Once again we assume that $\boldsymbol{\omega}$ has been set and thus we just have to account for the uncertainty in $\boldsymbol{\theta}$. We detail a Monte Carlo sampling strategy for propagating these uncertainties as shown in Algorithm~\ref{alg:mc_rakes}. We define $\boldsymbol{\mu}_{\boldsymbol{\theta}  } \in \mathbb{R}^{N}$ to be the mean rake position (which we can assume corresponds to the ones provided in Table 1 in part I \cite{seshadri2019a}) and $\boldsymbol{\varSigma}_{\boldsymbol{\theta}  } \in \mathbb{R}^{N \times N }$ to be a covariance matrix corresponding to the uncertainties in rake positions. We generate a series of $L$ random vectors $\vh_{i}$ where $i=1, \ldots, L$ that have these mean and covariance values. Then, we compute $\mA_{i}$ for each $\vh_{i}$ and proceed to solving for $\mX_{i}$.

\begin{algorithm}
\caption{Strategy for propagating rake uncertainties.}
\label{alg:mc_rakes}
\begin{algorithmic}[1]
\STATE{Set $\boldsymbol{\omega}=\left(\omega_1, \omega_2 \right)$, $\boldsymbol{\mu}_{\boldsymbol{\theta}  }$ and $\boldsymbol{\varSigma}_{\boldsymbol{\theta}  }$.}
\STATE{Set $L=50000$}
\STATE{Set $\boldsymbol{\lambda} = \left(0.0001, 0.001, 0.1, 10 \right)$}
\STATE{Generate random vectors $\mH = \text{rand}  \left(  \boldsymbol{\mu}_{\boldsymbol{\theta}  } ,  \boldsymbol{\varSigma}_{\boldsymbol{\theta}  } , L \right)$.}
\FOR{$i=1:L$}
\STATE{Compute $\mA_{i} = \mA \left( \boldsymbol{\omega} , \vh_{i}^{T}    \right)$}
\STATE{Solve $\hat{\mX}_{i}={\text{argmin}}\left\Vert \mA_{i}\mX-\mB\right\Vert _{2}^{2}$}
\WHILE{$\left\Vert \hat{\mX}_{i} \right\Vert _{2}\geq\beta$}
\STATE{$\lambda_i = \boldsymbol{\lambda}\left( i \right)$}
\STATE{Solve $\hat{\mX}_{i}={\text{argmin}}\left\Vert \mA_{i}\mX-\mB\right\Vert _{2}^{2} + \left\Vert \lambda_{i} \mX \right\Vert _{2}^{2}$  }
\ENDWHILE
\STATE{$\mathcal{X}(:, :, i) =\hat{\mX}_{i}$}
\ENDFOR
\RETURN $\mathcal{X}$ 
\end{algorithmic} \end{algorithm}

In a nutshell, Algorithm \ref{alg:mc_rakes} simply perturbs $\mA$ to account for the uncertainty in the rake positions; the regularized least squares approach from part I is utilized in lines 8 to 11 of this algorithm. The argument that is returned from this algorithm is a tensor $\mathcal{X} \in \mathbb{R}^{\left(2k+1 \right) \times M \times L}$ where, as before, the dimension $\left(2k + 1 \right)$ represents the number of coefficients corresponding to the $k$ harmonics. Each \emph{slice} of this tensor is the matrix $\hat{\mX}_{i}$ for $i=1, \ldots, L$. These slices can then be pre-multiplied by
\begin{equation}
\mA_{prediction} = \mA \left(\boldsymbol{\omega}, \boldsymbol{\theta}_{prediction} \right)
\end{equation}
with $\boldsymbol{\theta}_{prediction} \in \mathbb{R}^{P}$ discretizing the circumference (0$^\circ$ to 360$^\circ$) into $P$ elements, where $P$ is a large number. Predictive mean and variance values for $\mA_{prediction} \hat{\mX}_{i}$ can then be computed.

\begin{figure}
\centering
\includegraphics[width=11.0cm]{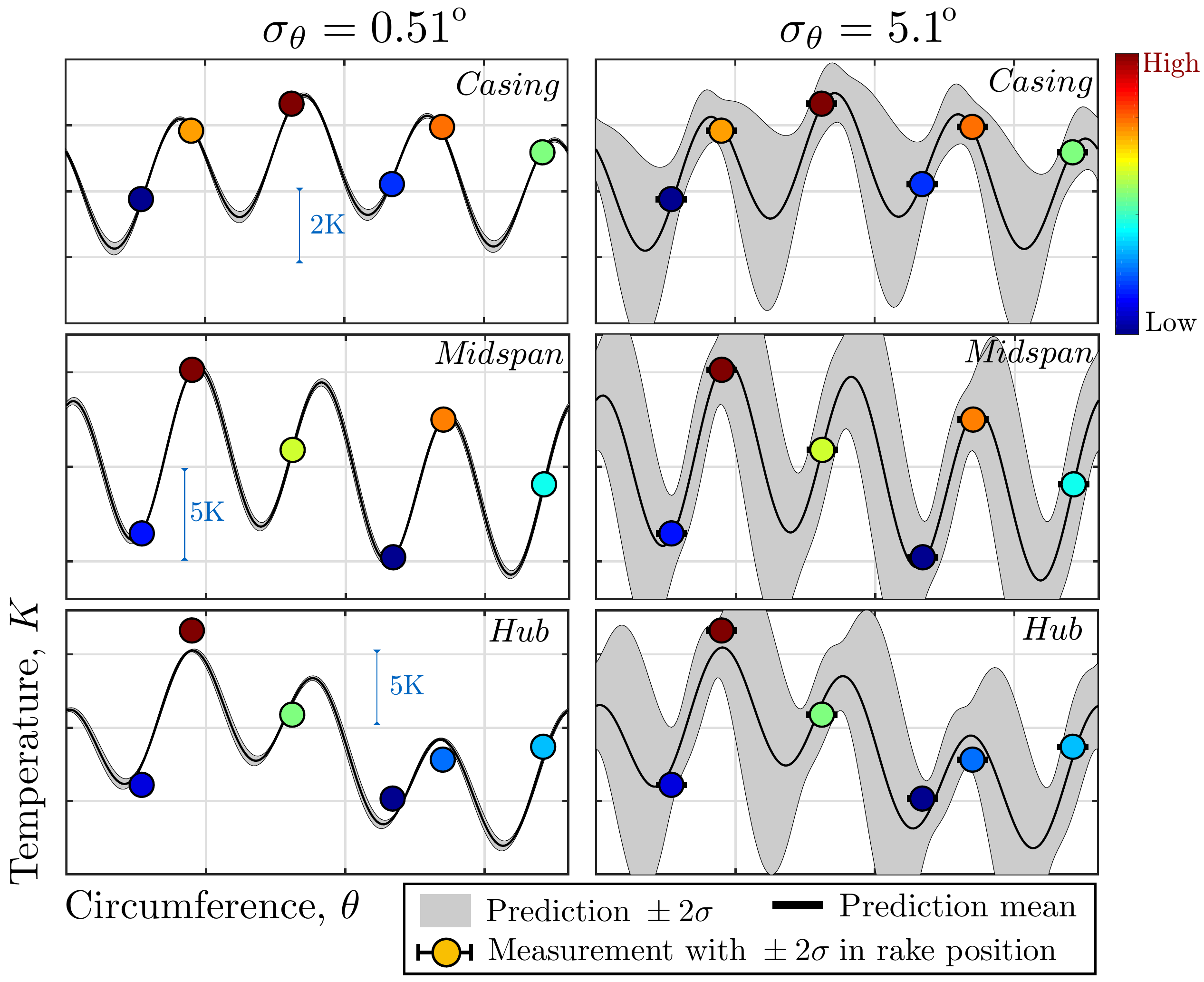}
\caption{Predictive mean and variance contours for uncorrelated rake placements with $\sigma_{\theta}=0.51^{\circ}$ (left) and $\sigma_{\theta}=5.1^{\circ}$ (right).}
\label{fig:uq5}
\end{figure}

In Fig.~\ref{fig:uq5} we plot the predictive mean and variance values for the same extract of Engine A considered previously. We set $\boldsymbol{\mu}_{\boldsymbol{\theta}}$ to correspond to the rake positions associated with Engine A (see Table 1 in \cite{seshadri2019a}) and $\boldsymbol{\varSigma}_{\boldsymbol{\theta}} = \sigma_{\theta}^2 \mI$, where we study two cases: $\sigma_{\theta} = 0.51^\circ$ and $\sigma_{\theta} = 5.1^\circ$. These correspond to an uncertainty in rake positions of $\pm 1^{\circ}$ and $\pm 10^{\circ}$ respectively, both at a confidence level of 95$\%$. While uncertainties in rake positions of the order of a few degrees may seem a possibility, a $\pm 10^{\circ}$ variation is unlikely. Our motivation in propagating these two rake uncertainties is to show that when approximating a signal with low frequency harmonics, such as $\boldsymbol{\omega} = (1, 4)$, a large (and unrealistic) variation in rake positions is required to induce a noticeable increase in the predictive variance. We remark that that the construction of $\boldsymbol{\varSigma}_{\boldsymbol{\theta}}$ implies that there are no correlations between the uncertainties in the $N$ rake positions.

\section{Re-visiting sampling uncertainty}
\label{sec:resultsB}
We opened this paper with a discussion on sampling uncertainty and remarked that its widely adopted definition falls short of quantitatively articulating what sampling uncertainty truly is. That said, the motivation for having a \emph{sampling uncertainty metric} is clear: we wish to ascertain whether the spatial variation of a flow quantity can be estimated with a finite number of circumferentially positioned rakes and their radial probes. However, we argue that such a metric cannot be independent of the bias and precision of its composite measurements. 

\subsection{Assumed temperature variations}
\begin{figure}
\centering
\includegraphics[width=11.0cm]{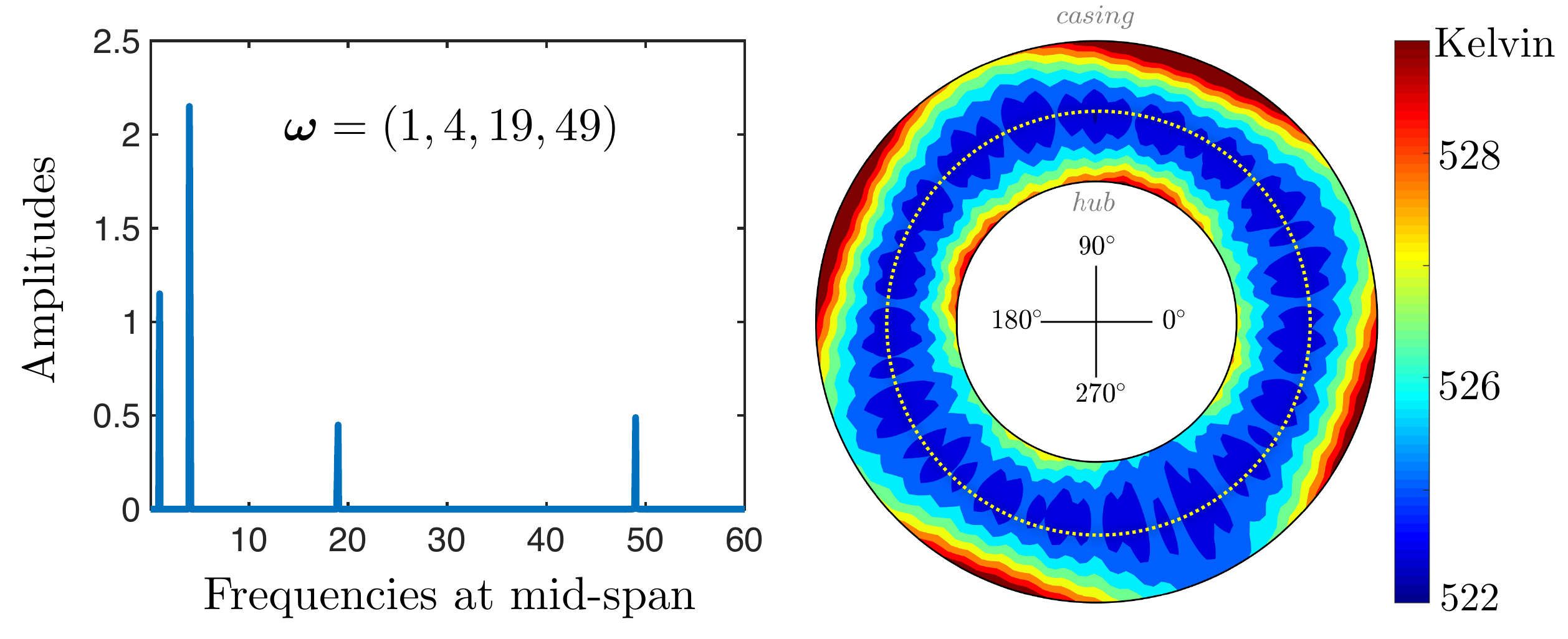}
\caption{An assumed temperature spatial profile with four harmonics.}
\label{fig:assumed_1}
\end{figure}

\begin{figure}
\centering
\includegraphics[width=11.0cm]{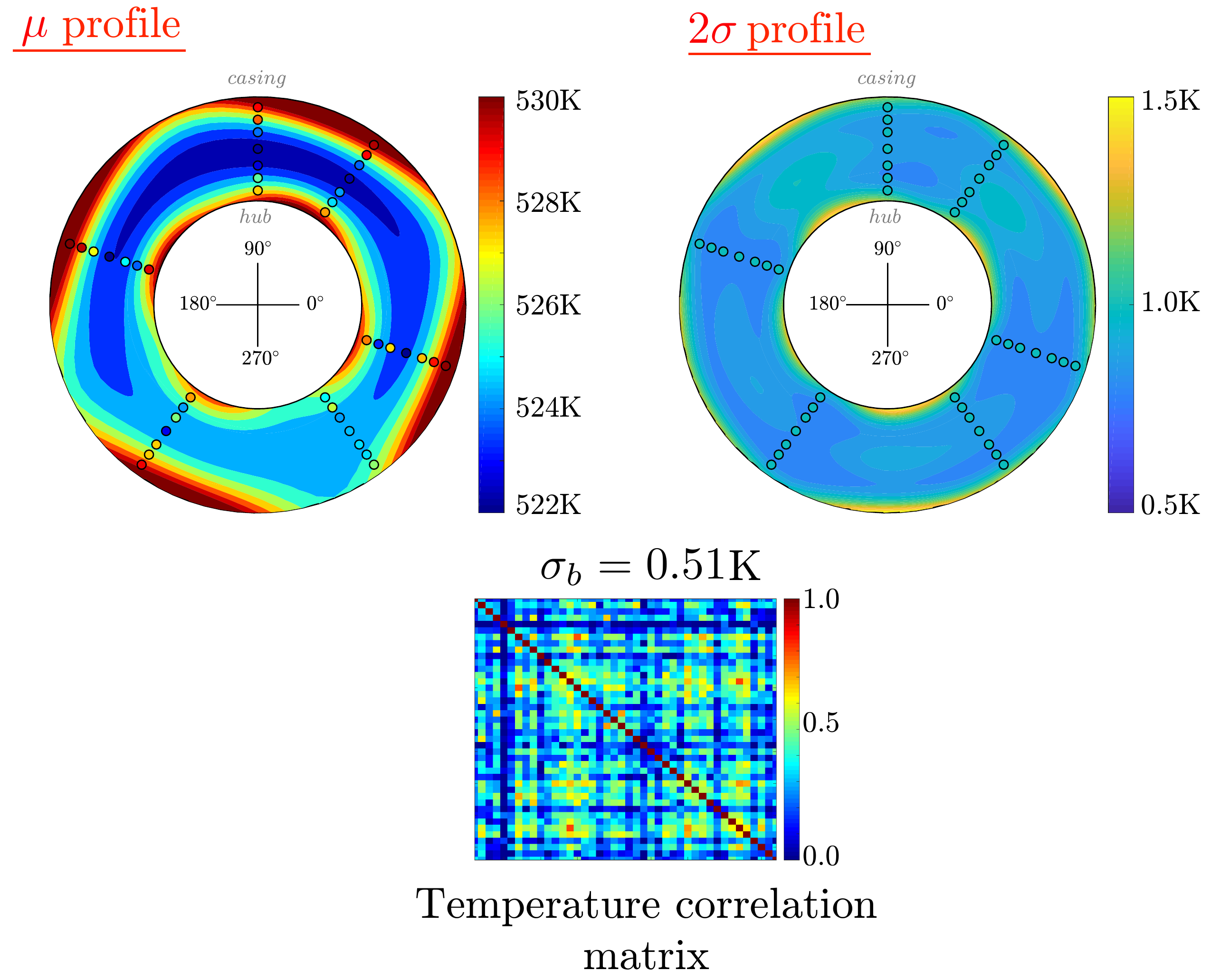}
\caption{Mean and $2\sigma$ profiles in the temperature for $\sigma_{b}=0.51$K and the shown correlation matrix.}
\label{fig:assumed_2}
\end{figure}

Recall the assumed temperature profile considered in part I, shown again in Fig.~\ref{fig:assumed_1}. This temperature profile has a mean temperature of 526.20 K and is comprised of harmonics $\boldsymbol{\omega} = \left(1, 4, 19, 49 \right)$ with known amplitudes and phases. Let us assume that each temperature measurement can be characterized by a Gaussian distribution with $\sigma_{b} = 0.51$K, just as before, and a certain correlation matrix. We sample the spatial field in Fig.~\ref{fig:assumed_1} at the rake positions corresponding to Engine A, i.e., $\boldsymbol{\theta} = (54^\circ, 90^\circ, 162^\circ, 234^\circ, 270^\circ, 342^\circ)$, and obtain spatial representations of the mean and uncertainty in temperature, as shown in Fig.~\ref{fig:assumed_2}. Computing the expectation of both these fields yields an average of $525.85 \pm 0.90$ K. 

In contrast, in Table \ref{table:uqstudy2}, we aggregate the formulas for the measurement and sampling uncertainty using the standard root-mean-square method. The sampling uncertainty is given by
\begin{align}
\begin{split}
\text{Sampling uncertainty} &=\sqrt{\frac{\sum_{i=1}^{K}\left(T_{i}-525.975\right)^2}{42-1}} \\
&=\pm 2.371\text{K}
\end{split}
\end{align}
where the $K=NM=42$ corresponds to the total number of measurements. The temperature value of 525.975 K is the numeric average of the temperature based on the values at these 42 spatial locations. It is readily apparent that the uncertainty confidence intervals predicted via the standard root-mean-square approach are almost three times larger than our value of $0.90$K. 

\begin{table}
\begin{center}
\caption{Uncertainty budget.}
\begin{tabular}{|c|c|}
  \hline		
 Uncertainty & Value \\ \hline
 \hline
Measurement uncertainty & $\pm 1$K \\
Sampling uncertainty &  $\pm 2.371$K  \\
Total uncertainty & $\pm 2.573$K \\
  \hline  
\end{tabular}
\label{table:uqstudy2}
\end{center}
\end{table}
\section*{Conclusions}
In this paper, we have developed a theoretical framework for propagating uncertainties into our supervised learning model from \cite{seshadri2019a}. Our results on engine data show that the uncertainties in individual temperature measurements (and the correlations therein) are more important when compared with the uncertainties associated with the position of the rakes.

One motivating idea behind our study was to re-evaluate the definition of sampling uncertainty. We demonstrate that such a metric---in the absence of any assumed spatial variation---falls short of achieving its objective of informing an engineer whether he / she needs to fit more rakes at a measurement plane. Our analysis showed that the bounds obtained from the widely adopted sampling uncertainty metric are pessimistic and far too large, which may lead to an erroneous interpretation of the overall uncertainties in engine tests. 

To address these shortcomings we offered two new metrics in this paper: one for quantifying the \emph{spatial sampling uncertainty} and another for characterizing \emph{measurement imprecision uncertainty}. 

\section*{Acknowledgments}
The authors are grateful to Ra\'{u}l V\'{a}zquez D\'{i}az (Rolls-Royce). This work was funded by Rolls-Royce plc; the authors are grateful to Rolls-Royce for permission to publish this paper.

\bibliography{references}
\bibliographystyle{asmems4}

\end{document}